\documentclass[aps,prb,twocolumn,amsmath,amssymb,amsfonts]{revtex4}
\pdfoutput=1

\usepackage{wasysym}

\usepackage{graphicx}
\usepackage{subfigure}

\newcommand{\bra}[1]{\left< #1\right|} 
\newcommand{\ket}[1]{\left|#1 \right>}

\newcommand{\avg}[1]{\langle #1 \rangle}

\newcommand{\h}[1]{{#1}^{\dagger}} 
\newcommand{\cc}[1]{{#1}^{*}}
\newcommand{\cb}[1]{\bar{#1}}

\newcommand{\up}{\uparrow}
\newcommand{\down}{\downarrow}

\newcommand{\del} {\partial}

\newcommand{\FeCrAs}{{\rm FeCrAs}}

\begin{document}

\title{Hidden spin liquid in an antiferromagnet: Applications to $\FeCrAs$}
\author{Jeffrey G. Rau}
\affiliation{Department of Physics, University of Toronto, Toronto, Ontario M5S 1A7, Canada}
\author{Hae-Young Kee}
\affiliation{Department of Physics, University of Toronto, Toronto, Ontario M5S 1A7, Canada}

\date{\today}

\begin{abstract}
The recently studied material FeCrAs exhibits a surprising combination of experimental signatures,
 with metallic, Fermi liquid like specific heat but resistivity showing strong non-metallic character.
The $\rm Cr$ sublattice posseses local magnetic moments, in the form of stacked (distorted) Kagome lattices.
Despite the high degree of magnetic frustration, anti-ferromagnetic order develops below $T_N \sim 125K$ suggesting
the non-magnetic $\rm Fe$ sublattice may play a role in stabilizing the ordering.
 From the material properties we propose a microscopic Hamiltonian for the low energy degrees of freedom, including
the non-magnetic $\rm Fe$ sublattice, and study its properties using slave-rotor mean field theory. Using this approach we  find a spin liquid phase on the $\rm Fe$ sublattice, which survives even in the presence of the magnetic $\rm Cr$ sublattice.
 Finally, we suggest that the features of FeCrAs can be qualitatively explained by 
critical fluctuations in the non-magnetic sublattice Fe
due to proximity to a metal-insulator transition. 
\end{abstract}

\pacs{}

\maketitle
\section{Introduction}

The ubiquity of Landau's Fermi liquid is a testament to
universality in the solid state. As such,
departures from these classic experimental signatures
in metallic systems act as a guide to novel and interesting physics.
Non-Fermi liquid behaviour appears in many strongly correlated materials
such as unconventional superconductors\cite{cuprate1,feas1,organics1}, heavy fermion materials\cite{stewart1,stewart2} and
near quantum phase transitions\cite{qcp1,qcp2}. Some routes to realize this behaviour include
coupling itinerant electronic systems to localized magnetic moments and through intermediate
to strong electron-electron interactions. These
mechanisms can give rise to characteristics and experimental
signatures that do not fit neatly in the Fermi liquid paradigm.

\begin{figure}
\subfigure[\ View along {$c$} axis]{
\begin{picture}(105,100)(0,0)
\includegraphics[scale=0.07]{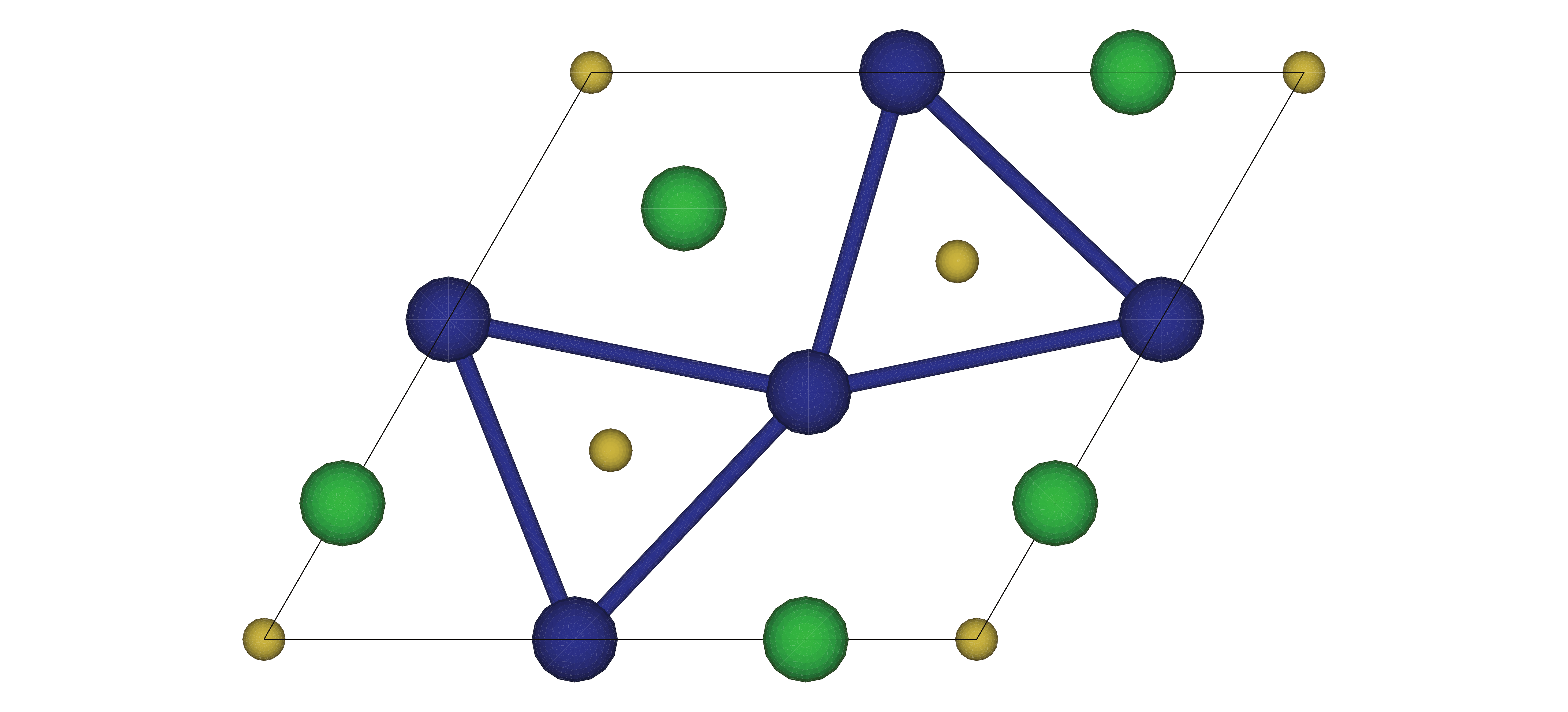}
  \put(-25,28){Cr}
  \put(-32,15){Fe}
  \put(-40,2){As}
\end{picture}}
\subfigure[\ Tilted view]{\includegraphics[scale=0.07]{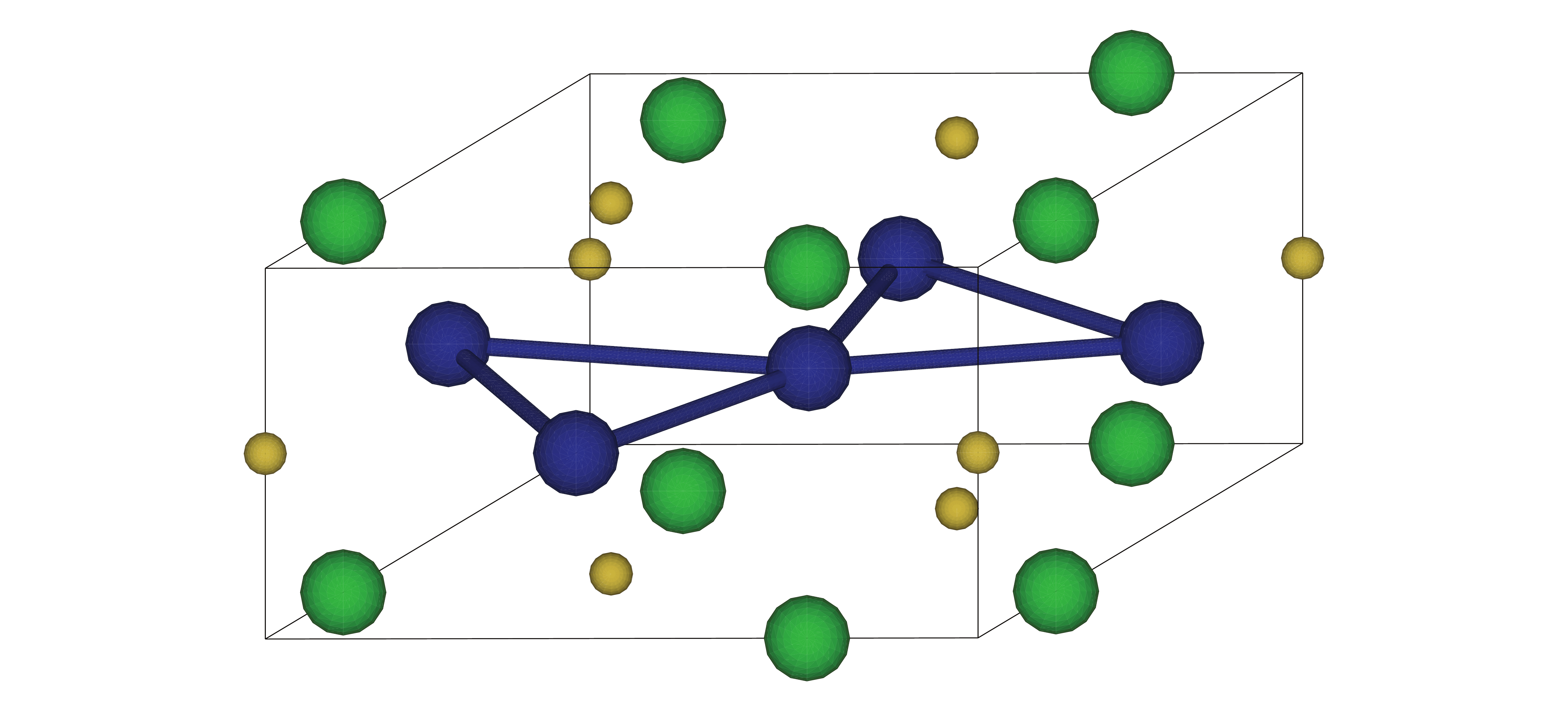}}
\caption{ \label{unit-cell}(Color Online) FeCrAs unit cell viewed along the $c$ axis (left) and tilted away by $70^{\circ}$ (right). Fe atoms are
indicated in green, Cr in blue and As in yellow.}
\end{figure}

\begin{figure}
\subfigure[\ Trimer Lattice]{\includegraphics[scale=0.065]{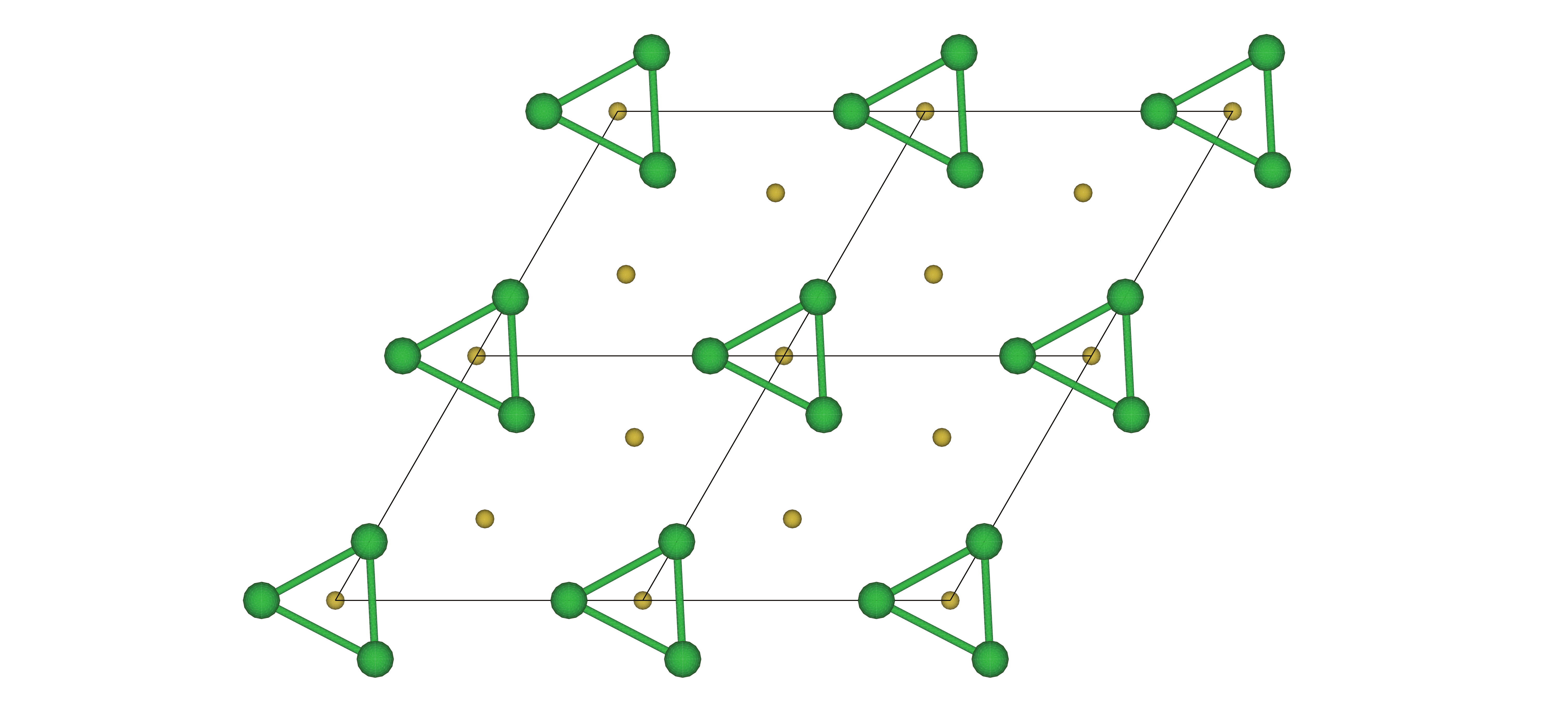}}
\subfigure[\ Relation to Cr sublattice]{\includegraphics[scale=0.065]{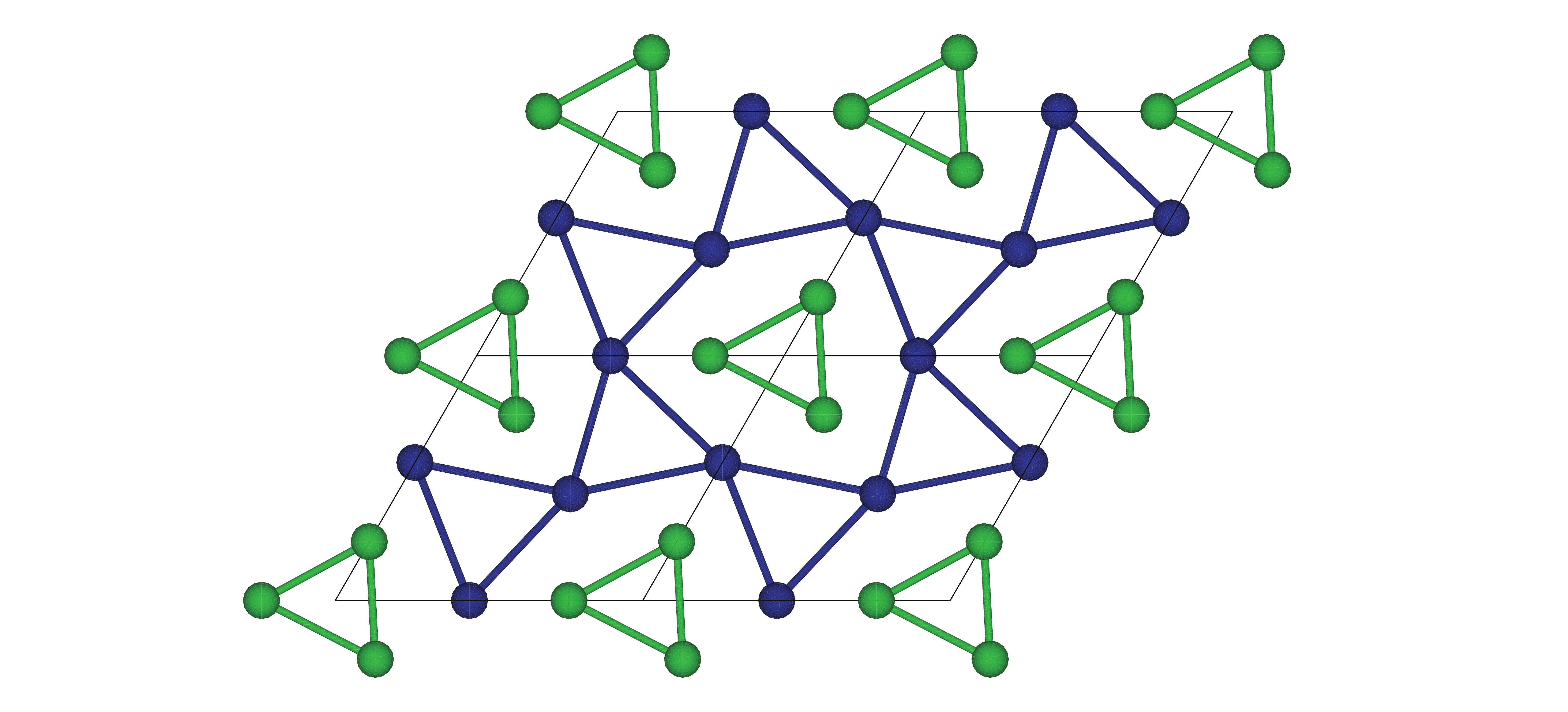}}
\caption{(Color Online) (a) The Fe sublattice with trimers shown explicitly.  (b) The local environment of the trimer, with
respect to the Cr sublattice.\label{trimer-lattice}}
\end{figure}

A recently re-examined compound, $\FeCrAs$\cite{fruchart1,fruchart2,julian}, provides a direct example
of a material that does not fit completely within Fermi liquid theory and
combines aspects of the mechanisms discussed above.
The unit cell of $\FeCrAs$ shown in Fig. \ref{unit-cell} shows $\rm Cr$ and $\rm Fe$ form alternating two dimensional lattices along
what we will denote the $c$ axis. $\rm Cr$ forms layers with the structure of 
a distorted Kagome lattice where the $\rm Cr-Cr$ distances are approximately
constant. The $\rm Fe$ layers have a more complicated structure, forming a triangular lattice of three atom units which we will call trimers as shown in Fig. \ref{trimer-lattice}.
The $\rm As$ is interspersed throughout
both the $\rm Fe$ and $\rm Cr$ layers, as well as in between.
Most of the known experimental data is nicely presented in Wu et al\cite{julian}, which we summarize below.

The specific heat exhibits Fermi liquid behaviour at low temperatures, i.e. $C \sim \gamma T$
where the slope $\gamma$ is sample dependent\cite{julian-synth}. The measured linear
range is roughly $T \sim 3 K - 10 K$.
Resistivity measurements show insulating behaviour at low and high temperatures. In plane ($ab$) and out of 
plane ($c$) resistivities are of the same order over the entire temperature range considered. The resistivity monotonically decreases (that is $d\rho/dT<0$)
as $T$ is raised from $5K$ up to $\sim 800K$ except for a small peak in the $c$ axis resistivity around $T \sim 125 K$.
A low temperature power law $\rho \sim \rho_0 - A T^{\alpha}$ is observed for $T \sim 80mK - 5 K$ in both $ab$ planes and $c$ axis resistivity with $\alpha \sim 0.6-0.7$

There is a peak in the susceptibility at $T_N \sim 125 K$ indicating a magnetic transition with a lack of hysteresis
pointing to antiferromagnetic ordering. Below $T_N$ the susceptibility is anisotropic, differing between the
$ab$ plane and the $c$ axis.
Elastic neutron scattering\cite{julian-neutron} done deep in the magnetic phase, at $T = 2.8 K$, is consistent
with the anti-ferromagnetic order inferred from the susceptibility,
signaling an ordering vector at $\vec{Q} = (\frac{1}{3},\frac{1}{3},0)$, indicating ferromagnetic (stacked) 
order along the c-axis, but with a tripled unit cell in the $ab$ plane. 
While measurements of the specific heat give a result consistent with a metallic Fermi liquid,
transport is unusual and deviates strongly from the classic Fermi liquid result for metals, while
being distinct from the expected result for strong insulators. Furthermore,
this material has a frustrated magnetic sublattice that nonetheless orders at low temperatures, while
the remaining sublattices show either small or no magnetic moment\cite{julian,julian-neutron}. The magnetic sublattice takes the form of a distorted
Kagome lattice, where even classical Heisenberg models fail to order magnetically for both the stacked\cite{stacked-kagome-1,stacked-kagome-2} and purely two dimensional cases\cite{kagome-1,kagome-2,kagome-3,kagome-4,kagome-5}.
Very few experiments have been carried out on $\FeCrAs$, so theoretical models are not completely restricted.
Regardless, there are a number of questions that need to be addressed, such as the nature of the stabilization
of magnetic order, the cause of the very different thermodynamic and transport signals and the
role of the non-magnetic sublattice. 

The nature of the magnetic order has been recently
addressed by Redpath et al\cite{hopkinson}, where a minimal model was proposed which suffices to explain the experimentally
observed stabilization of a particular magnetic ordering vector.
However, transport and thermodynamic behavior remains to be explained along with the role of the non-magnetic, ${\rm Fe}$ sublattice.
In this paper, we elaborate a microscopic route to an effective model for the compound $\FeCrAs$,
taking into account the ${\rm Fe}$ sublattice, and present
a scenario to address the incongruities between the conflicting metallic, Fermi liquid specific heat and insulating like
transport signals. This model consists of interacting electrons of the non-magnetic sublattice coupled
to magnetic moments of the magnetic, ${\rm Cr}$ sublattice.  Here
we do not address the detailed nature of the moments themselves, treating them classically\cite{hopkinson},
with the non-Fermi liquid physics arising from strong charge fluctuations occuring at
intermediate Hubbard coupling in the ${\rm Fe}$ sublattice. In this picture we are
excluding any Kondo physics, a view supported by the experimental results.
Our emphasis is on the interplay between strong charge fluctuations near the metal-insulator transition
on the $\rm Fe$ sublattice and the magnetic order of the $\rm Cr$ sublattice.
For this we turn to the slave-rotor method which allows access to the intermediate coupling regime 
and metal-insulator transition.

The structure of the paper is as follows: in Section \ref{effham} we present
an argument to pass from the atomic limit through to an effective model of the electronic degrees 
of freedom in $\FeCrAs$. In Section \ref{model} we discuss the localized moments and magnetic interactions
and we present the effective Hamiltonian relevant for $\FeCrAs$. We proceed to review the slave-rotor method
in Section \ref{slave-rotors} and the assumptions and implementation of our mean field theory in Section \ref{mean-field-theory}. In
Section \ref{discussion} we comment on the application of our results to $\FeCrAs$ and summarize conclusions in Section \ref{conclusions}.
\begin{figure}
\subfigure[\ Tetrahedral]{\includegraphics[scale=0.07]{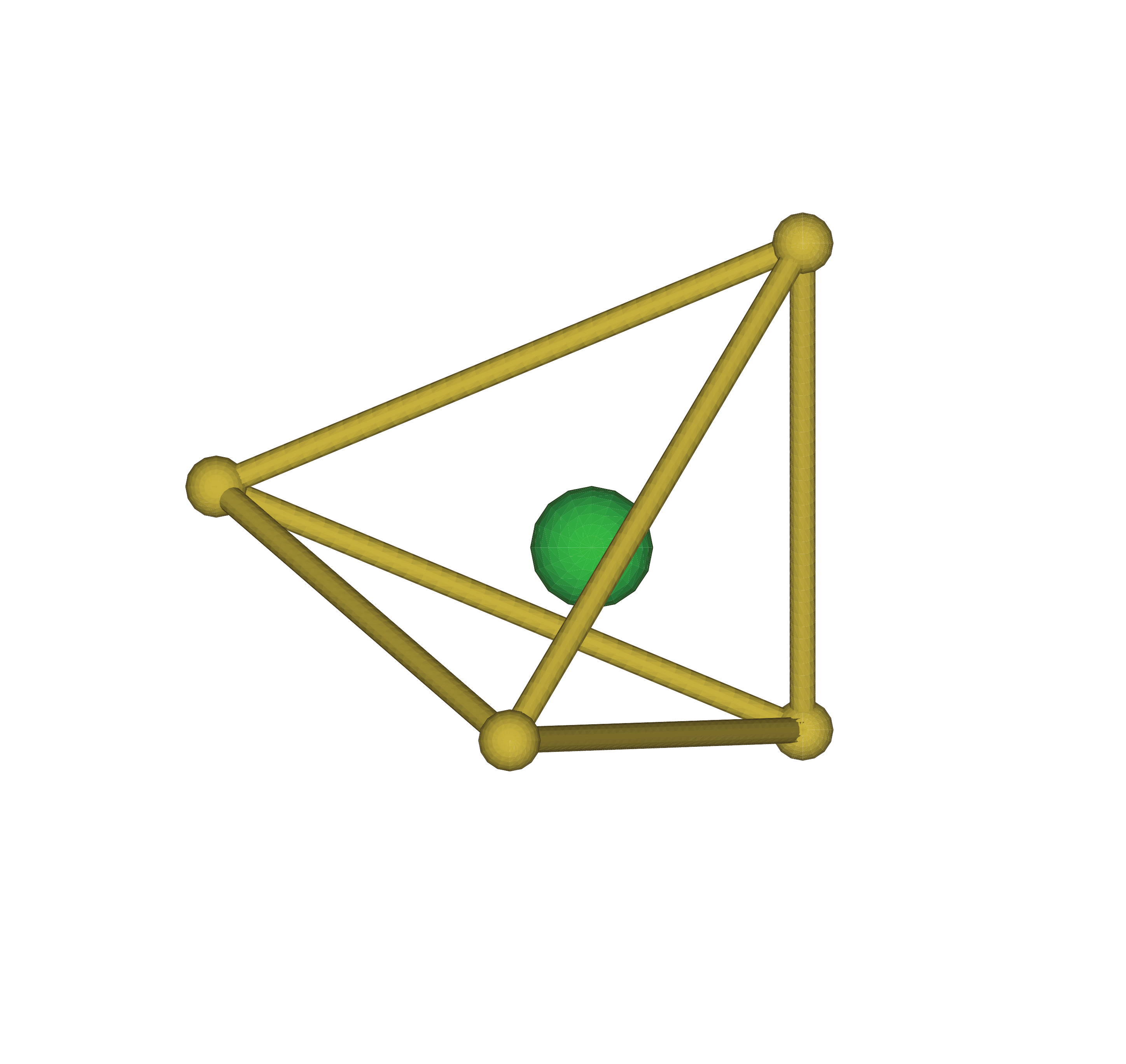}}
\subfigure[\ Pyramidal]{\includegraphics[scale=0.07]{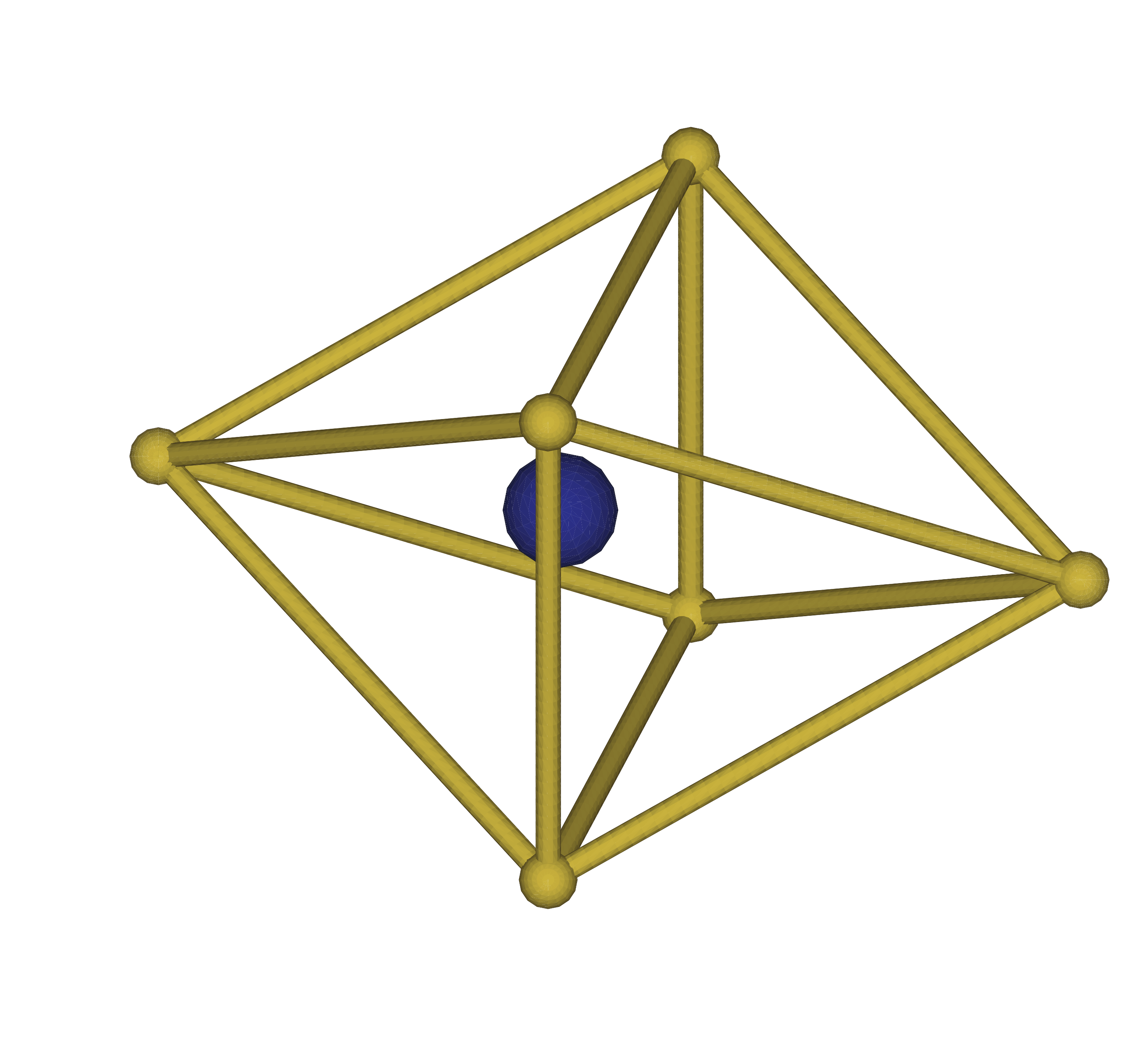}}
\subfigure[\ Trimer]{\includegraphics[scale=0.08]{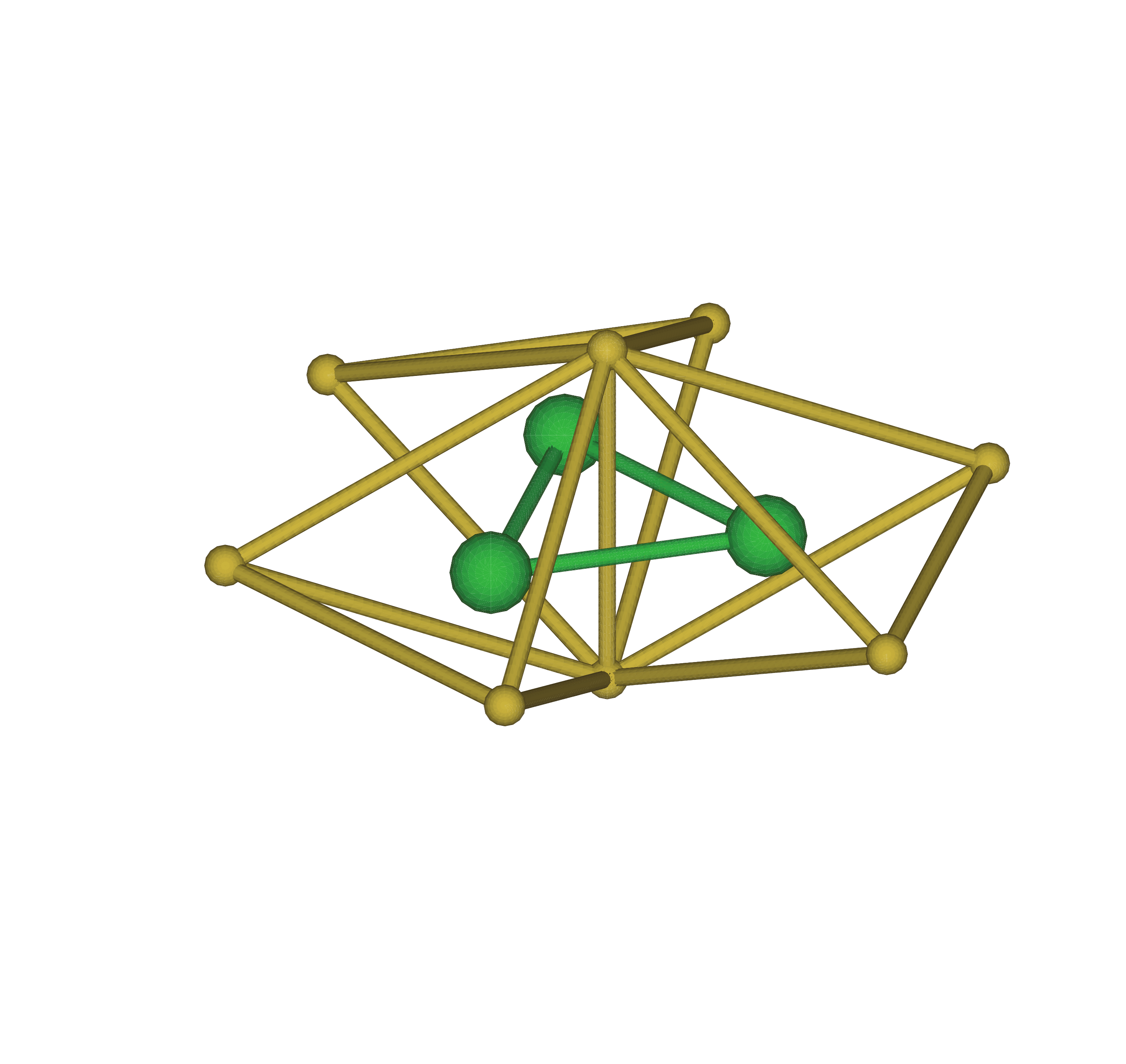}}
\caption{(Color Online) The local environments of the $\rm Fe$ (a), $\rm Cr$ (b), and trimer (c). The 
Fe atom is tetrahedrally coordinated and the Cr atom is approximately octahedrally coordinated. The
trimer is surrounded by three tetrahedra that share a common axis.\label{local-geometry}}
\end{figure}
\section{Effective Hamiltonian}
\label{effham}

\subsection{Local environments and spin states} 

\begin{figure}
\subfigure[\ Effect of the tetrahedral crystal field]{
\setlength{\unitlength}{0.6mm}
\begin{picture}(50,60)(20,20)
  \put(20,50){\line(1,0){20}}
  \put(20,51){\line(1,0){20}}
  \put(20,52){\line(1,0){20}}
  \put(20,53){\line(1,0){20}}
  \put(20,54){\line(1,0){20}}
  \put(25,43){\footnotesize $l=2$}
  \put(33,65){{\footnotesize $xz$}}
  \put(33,70){{\footnotesize $yz$}}
  \put(33,75){{\footnotesize $xy$}}
  \put(23,27){{\footnotesize $x^2-y^2$}}
  \put(23,32){{\footnotesize $3z^2-r^2$}}
  \put(40,70){\line(1,0){20}}
  \put(40,71){\line(1,0){20}}
  \put(40,72){\line(1,0){20}}
  \put(65,71){$t_{2}$}
  \put(50,50){\vector(0,1){16}}
  \put(55,50){$\Delta_{T_d}$}
  \put(50,50){\vector(0,-1){16}}
  \put(40,30){\line(1,0){20}}
  \put(40,31){\line(1,0){20}}
  \put(65,31){$e$}
\end{picture}}
\hspace{0.06\textwidth}
\subfigure[\ Low spin state for Fe$^{3+}$]{
\setlength{\unitlength}{0.6mm}
\begin{picture}(50,60)(20,20)
  \put(50,70){\large $\uparrow$}
  \put(40,70){\line(1,0){20}}
  \put(40,71){\line(1,0){20}}
  \put(40,72){\line(1,0){20}}
  \put(65,71){$t_2$}
  \put(33,65){{\footnotesize $xz$}}
  \put(33,70){{\footnotesize $yz$}}
  \put(33,75){{\footnotesize $xy$}}
  \put(23,27){{\footnotesize $x^2-y^2$}}
  \put(23,32){{\footnotesize $3z^2-r^2$}}
  \put(41,29.5){\large  $\uparrow$}  
  \put(46,29.5){\large $\downarrow$}
  \put(51,29.5){\large  $\uparrow$}
  \put(56,29.5){\large $\downarrow$}
  \put(40,30){\line(1,0){20}}
  \put(40,31){\line(1,0){20}}
  \put(65,31){$e$}
\end{picture}
}
\caption{\label{fe-cf} The effect of the crystal field on the Fe$^{3+}$ ion. The local symmetry 
about this site is tetrahedral, that is the group $T_d$.}
\end{figure}
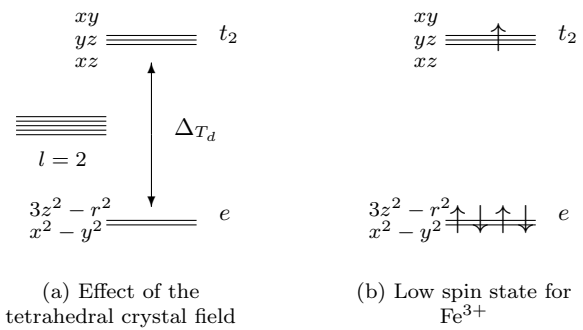
Considering the common oxidation states of $\rm Fe$
we will take (in the atomic limit) $\rm Fe^{3+}$ 
as a starting point. This leaves the valence configuration being
${\rm 3d}^5$ for ${\rm Fe}^{3+}$.
Following the experimental data, we assume that the Fe atoms are in a low spin state,
due to the lack of a detectable magnetic moment\cite{julian-neutron}.  The tetrahedral 
arrangement of the As atoms about each Fe atom, shown in Fig. \ref{local-geometry} (a), implies a crystal field
producing a splitting of the Fe $d$ levels, shown in Fig. \ref{fe-cf} (a), with a pair of low-lying $e$ levels and a three-fold degenerate 
set  of $t_{2}$ levels, separated by the crystal field gap.
Assuming the low spin case is relevant the two $e$ states are filled
and there is a single electron in the $t_{2}$ triplet shown in Fig. \ref{fe-cf} (b).

\begin{figure}[tp]
\subfigure[Effect of the tetragonal crystal field]{
\setlength{\unitlength}{0.6mm}
\begin{picture}(70,60)(20,20)
  \put(20,50){\line(1,0){20}}
  \put(20,51){\line(1,0){20}}
  \put(20,52){\line(1,0){20}}
  \put(20,53){\line(1,0){20}}
  \put(20,54){\line(1,0){20}}
  \put(25,43){\footnotesize $l=2$}
  \put(40,30){\line(1,0){20}}
  \put(40,31){\line(1,0){20}}
  \put(40,32){\line(1,0){20}}
  \put(30,31){$t_{2g}$}
  \put(63,35){\line(1,0){20}}
  \put(63,36){\line(1,0){20}}
  \put(85,35){$e$}
  \put(63,25){\line(1,0){20}}
  \put(85,25){$b_2$}
  \put(50,50){\vector(0,1){16}}
  \put(52,50){$\Delta_{O_h}$}
  \put(50,50){\vector(0,-1){16}}
  \put(40,70){\line(1,0){20}}
  \put(40,71){\line(1,0){20}}
  \put(30,71){$e_{g}$}
  \put(63,65){\line(1,0){20}}
  \put(85,65){$b_1$}
  \put(63,75){\line(1,0){20}}
  \put(85,75){$a_1$}
  \put(73,70){\vector(0,1){4}}
  \put(73,70){\vector(0,-1){4}}
  \put(75,69){$\Delta^{+}_{D_4}$}
  \put(73,30){\vector(0,-1){4}}
  \put(73,30){\vector(0,1){4}}
  \put(75,29){$\Delta^{-}_{D_4}$}
  \put(73,50){\vector(0,-1){10}}
  \put(73,50){\vector(0,1){10}}
  \put(75,49){$\Delta^{0}_{D_4}$}
\end{picture}}
\hspace{0.02\textwidth}
\subfigure[Low spin state for Cr$^{2+}$]{
\setlength{\unitlength}{0.6mm}
\begin{picture}(50,60)(50,20)
  \put(63,35){\line(1,0){20}}
  \put(63,36){\line(1,0){20}}
  \put(85,35){$e$}
  \put(63,25){\line(1,0){20}}
  \put(85,25){$b_2$}
  \put(63,65){\line(1,0){20}}
  \put(85,65){$b_1$}
  \put(63,75){\line(1,0){20}}
  \put(85,75){$a_1$}
  \put(78,34){\large $\uparrow$}
  \put(78,24){\large $\uparrow$}
  \put(68,24){\large $\downarrow$}
  \put(68,34){\large $\uparrow$}
\end{picture}
}\caption{\label{cr-cf} The effect of crystal fields on the Cr$^{2+}$ ion.
The Cr ion is biased towards one direction of the octahedron. This reduces the symmetry to the tetragonal
group $D_4$.
}
\end{figure}
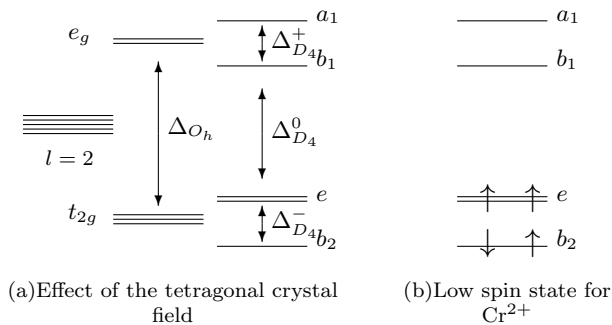

The case of $\rm Cr$ is more complicated, as the experiments do not single out a more probable spin state.
For $\rm Cr$ we will take a ionic charge of ${\rm Cr}^{2+}$, giving a valence configuration of $3d^4$.
A distorted octahedral environment, shown in Fig. \ref{local-geometry} (b) reduces the symmetry about this
site to $D_4$, with crystal field splittings shown in Fig. \ref{cr-cf} (a).
The low spin state(shown in Fig. \ref{cr-cf} (b)) can still yield a $S=1$ spin moment, while 
the high spin state has an $S=2$ moment. In the low spin case hopping onto the Cr$^{2+}$ will be suppressed by
the orbital repulsion, while for the high spin case the crystal field energy will give a further supression.

We would like to emphasize that the arguments and models presented here are under-constrained
by both experimental results and first principles electronic structure calculations\cite{abinitio}. Due to this limitation
the precise details could fail quantitatively, but the subsequent effective Hamiltonian appears to be robust to
a variety of ionic configuration changes in the underlying model. For example, the specific oxidation state we use for the $\rm Cr$ will be irrelevant to our
final discussion, as only the localized character is needed to capture the gross magnetic features\cite{hopkinson}.
With this in mind below we present a possible route from the microscopic Hamiltonian to the effective model.

\subsection{Iron-Chromium interactions}
Since the Fe-Cr distance is considerably smaller than the direct Fe-Fe distances outside the trimers
and the indirect Fe-As-Cr distance, we will consider interactions induced by Fe-Cr-Fe hopping
paths. First we define a simple model for an isolated Cr atom,
assuming a low spin state as shown in Fig. \ref{cr-cf} (b). The local Hamiltonian for the $\rm Cr$ $e$ doublet assumes
the natural form
\begin{equation}
  H_{{\rm Cr}} = \Delta \sum_{j\alpha} n_{j\alpha} + U \sum_{j\alpha} n_{j\alpha \uparrow} n_{j\alpha \downarrow} - J\sum_j\left(\sum_{\alpha} \vec{S}_{j\alpha}\right)^2,
\end{equation}
where $n_{j\alpha \sigma} = \h{e}_{j\alpha \sigma} e_{j\alpha \sigma}$ is the number operator for the state in the doublet $\alpha$. We have denoted the atomic potential
as $\Delta$, the intra-orbital repulsion as $U$ and the Hund's coupling as $J$. 
Allowing for hopping between $\rm Fe$ and $\rm Cr$, we include the term
\begin{equation}
  H_{{\rm Cr}-{\rm Fe}} = -\sum_{ij \sigma \alpha}
 t^{\alpha}_{ij} \h{d}_{i \sigma} e_{j\alpha \sigma} + {\rm h.c},
\end{equation}
where $i$ runs over the orbitals on sites connected to the Cr atom at site $j$ and $d$ is the annihilation operator 
on the Fe site $i$.
Integrating out all of the high energy degrees of freedom on the Cr atom,
the ground state for sufficiently large $J$ is given by the doubly occupied, $S=1$ triplet 
states which we denote by $\ket{t_{ja}}$ with $a=0,\pm$. 
Noting that $H_{\rm Cr-Fe}$ 
does not connect triplet to triplet states, as it changes the electron number on the Cr atom
we can formulate the effective Hamiltonian following
\begin{eqnarray}
  H_{\rm eff} &=& P_t H_{\rm Cr-Fe} \left(E_0 - H_{\rm Cr}\right)^{-1} \h{H}_{\rm Cr-Fe} P_t,
\end{eqnarray}
where $E_0$ is the triplet energy and $P_t =\sum_{ja} \ket{t_{ja}}\bra{t_{ja}}$ projects onto the triplet subspace.
Carrying out the expansion of the effective Hamiltonian, denoting
the spin operators on the $\rm Fe$ and $\rm Cr$ atoms as $\vec{S}_i$ and $\frac{1}{2}\vec{\sigma}_i$ respectively, our full effective Hamiltonian is given by
\begin{equation}
  H_{\rm eff} = -\sum_{ijl} \frac{1}{W}\left(\sum_\alpha t^{\alpha}_{ij} \cc{(t^{\alpha}_{lj})}\right)
\h{d}_{i}\left(1-\vec{\sigma} \cdot \vec{S_j}\right) d_{l},
\end{equation}
where 
\begin{equation}
  \frac{1}{W} = 
 \frac{ U+\frac{5 J}{2}}{\Delta^2 + \left(\frac{5 J}{4}\right)^2 + U(\frac{5J}{4}-\Delta)} > 0.
\end{equation}
For $i=l$ this represents an anti-ferromagnetic exchange between
the $\rm Fe$ and $\rm Cr$ atoms, while $i \neq l$ presents
a spin-dependent hopping term, non-zero for both intra-trimer and inter-trimer hopping paths.
Based on overlap of the orbital wavefunctions, 
we assume that the hopping $\sum_\alpha |t^\alpha_{ij}|^2$ is dominant for $i=l$ and independent of the orbitals,
leading to the effective exchange Hamiltonian
\begin{equation}
  H^{\rm exch}_{\rm eff} = \sum_{ij} \frac{1}{W}\left(\sum_\alpha |t^\alpha_{ij}|^2\right)
\h{d}_i \left( \sigma \cdot \vec{S}_j \right) d_i,
\end{equation}
up to a shift of the chemical potential. For future use, we will denote the effective exchange as
\begin{equation}
  J_K = \frac{1}{W} \sum_\alpha |t^\alpha_{ij}|^2 > 0.
\end{equation}
\subsection{Trimer Approximation for $\rm Fe$}
From the inter-atomic distances we expect that the Fe-Fe atoms in the
trimer structure are tightly coupled. We can take advantage of this by
grouping the degrees of freedom in the trimer into a single unit and formulating
the rest of our theory in terms of these variables. This process is similar to the
treatment of the pairs of molecules as the effective degrees of freedom, frequently employed
in studies of organic superconductors\cite{organics1}.
This assumes the energy
scales for interactions and inter-trimer hopping matrix elements are small relative to
the intra-trimer hoppings. In addition, we need that the tetrahedral
crystal field splitting be much larger than our intra-orbital interactions
(this is the low spin assumption) and both the intra- and inter-trimer hopping elements.

With this in mind, let us find the low energy degrees of freedom of a trimer, keeping only 
the three site cluster. Considering the $C_3$ symmetry, there are four possiblities for
the degeneracies of the two lowest levels, leading to either a half or quarter filled band
as the relevant states (assuming all gaps are large compared to the relevant energy scales).
Motivated by the Slater-Koster argument in Appendix A, we start with a model for the $xy$, $xz$ and $yz$
orbitals of the trimer:
\begin{eqnarray}
  H_{\rm trimer} &=& -\sum_{\avg{ij}} \h{d}_i 
\left(
\begin{tabular}{ccc}
$0$ & $t_{xz,yz}$ & 0 \\
$t_{xz,yz}$ & $0$ & 0 \\
$0$ & $0$ & $t_{xy,xy}$  
\end{tabular}
\right) d_j,
\end{eqnarray}
where $\h{d}_i = \left( \h{d}_{i,xz}\ \h{d}_{i,yz}\ \h{d}_{i,xy}\right)$ and $i,j=1,2,3$ are the sites in the trimer.
The $xy$ part is just a simple three site chain with a ground state at
$-2t_{xy,xy}$ and pair of excited levels at $t_{xy,xy}$. For the $xz$ and $yz$ orbitals
we can diagonalize $H_{\rm trimer}$ by changing the basis:
\begin{eqnarray}
  d_{i,+} &=& \frac{1}{\sqrt{2}}\left( d_{xz} + d_{yz}\right), \\ 
  d_{i,-} &=& \frac{1}{\sqrt{2}}\left( d_{xz} - d_{yz}\right).
\end{eqnarray}
This gives a pair of decoupled three site chains with hoppings $\pm t_{xz,yz}$ 
and thus energy levels of $\pm 2 t_{xz-yz}$ and  $\pm t_{xz,yz}$. When 
$\frac{1}{2}t_{xy,xy} < t_{xz,yz} < 2 t_{xy,xy}$ a single half-filled level in the trimer, symmetric under permutations of the three sites is realized.
Ignoring all of the other matrix elements in the complete hopping matrix, we find (see Appendix A)

\begin{eqnarray}
  t_{xz,yz} & \sim &\frac{1}{32}\left(7t_{\delta} - 16t_{\pi} + 9 t_{\sigma}\right), \\
  t_{xy,xy} & \sim &\frac{1}{64}\left(49 t_{\delta} - 12 t_{\pi} + 3 t_{\sigma}\right).
\end{eqnarray}

\begin{figure}[t]
\vspace{10pt}
\subfigure[Spectrum when $t_{xz,yz} > t_{xy,xy}$]{
\setlength{\unitlength}{0.5mm}
\begin{picture}(80,60)(30,20)
  \put(63,37){\line(1,0){20}}
  \put(63,38){\line(1,0){20}}
  \put(63,64){\line(1,0){20}}
  \put(63,65){\line(1,0){20}}
  \put(85,37){$\scriptstyle -t_{xz,yz}$}
  \put(63,25){\line(1,0){20}}
  \put(85,25){$\scriptstyle -2t_{xz,yz}$}
  \put(63,65){\line(1,0){20}}
  \put(85,64){$\scriptstyle t_{xz,yz}$}
  \put(63,75){\line(1,0){20}}
  \put(85,75){$\scriptstyle 2t_{xz,yz}$}
  \put(40,29){\line(1,0){20}}
  \put(25,60){$\scriptstyle t_{xy,xy}$}
  \put(40,60){\line(1,0){20}}
  \put(20,29){$\scriptstyle -2t_{xy,xy}$}
  \put(40,61){\line(1,0){20}}
  \put(47,28){$\downarrow$}
  \put(76,24){$\uparrow$}
  \put(68,24){$\downarrow$}
\end{picture}
}
\subfigure[Spectrum when $t_{xy,xy} > t_{xz,yz}$]{
\setlength{\unitlength}{0.5mm}
\begin{picture}(70,60)(30,20)
  \put(63,44){\line(1,0){20}}
  \put(63,45){\line(1,0){20}}
  \put(63,57){\line(1,0){20}}
  \put(63,58){\line(1,0){20}}
  \put(85,44){$\scriptstyle -t_{xz,yz}$}
  \put(63,35){\line(1,0){20}}
  \put(85,35){$\scriptstyle -2t_{xz,yz}$}
  \put(85,57){$\scriptstyle t_{xz,yz}$}
  \put(63,67){\line(1,0){20}}
  \put(85,67){$\scriptstyle 2t_{xz,yz}$}
  \put(40,29){\line(1,0){20}}
  \put(25,60){$\scriptstyle t_{xy,xy}$}
  \put(40,60){\line(1,0){20}}
  \put(20,29){$\scriptstyle -2t_{xy,xy}$}
  \put(45,28){$\downarrow$}
  \put(55,28){$\uparrow$}
  \put(72,34){$\uparrow$}
  \put(40,61){\line(1,0){20}}
\end{picture}
}
\caption{\label{trimer-spectrum-half} The two cases that
give rise to a half-filled band, with $t_{xy,xy} \approx t_{xz,yz}$. We can
either have a band with (a) $d_{xy}$ character or with (b) $d_{+}$ character.
}
\end{figure}
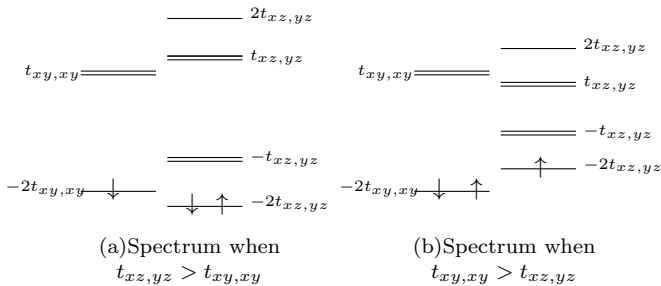

A more complete model would include mixing between all three orbitals but
the qualitative picture should remain the same. The naive choice of $t_{\delta}>t_{\pi}>t_{\sigma}$
seems to select the case of $t_{xy,xy}>t_{xz,yz}$ and thus 
the single occupied band has $d_{+}$ character (shown in Fig. \ref{trimer-spectrum-half}). Note the gap between
the occupied level and the half-filled level is given by
\begin{equation}
  \Delta_{\rm trimer} = 2|t_{xy,xy}-t_{xz,yz}|=\frac{5}{64}\left|7t_{\delta}+4t_{\pi} - 3 t_{\sigma}\right|,
\end{equation}
and must be a fairly large to ensure that the trimer approximation
is valid. We emphasize that regardless of the exact details of the these hoppings
the qualitative picture will remain, as long as this half-filling is found
such as for ${\rm Fe^{3+}}$. Other oxidation states will give 
a different relevant molecular orbital, possibly with extra orbital degeneracy, and
could change the picture presented here.

\subsection{Effective Hamiltonian and magnetic interactions}
\label{model}
The gross magnetic structure of
$\FeCrAs$ is captured by a classical model of localized moments interacting
with the $\rm Fe$ sublattice via a simple exchange\cite{hopkinson}. The utility of the classical model for the moments
is also supported by the fact that Kondo-like signatures
are absent in the experimental data. Futhermore, due to the uniform character
of the relevant trimer states and the assumption of equal exchanges between $\rm Cr$ and different $\rm Fe$ orbitals, the effective exchange between
the $\rm Cr$ and the trimers will be equal and between all of the $\rm Cr$
 in the surrounding hexagons in the layers above and below. 
 We thus take our Hamiltonian to have the form
\begin{widetext}
\begin{eqnarray}
  H =
-t \sum_{\avg{ij}\in {ab}}\sum_\sigma \h{c}_{i\sigma} c_{j\sigma}
-t' \sum_{\avg{ij}\in c}\sum_{ \sigma} \h{c}_{i\sigma} c_{j\sigma}+ U \sum_{i} n_{i\uparrow}n_{i\downarrow} 
 +\frac{J_K}{2}\sum_{i,a \in \hexagon_i} \left(\h{c}_i \vec{\sigma} c_i\right) \cdot \vec{S}_a
+ J_H \sum_{\avg{ab}} \vec{S}_a \cdot \vec{S}_b \label{full-model},
\end{eqnarray}
\end{widetext}
where the low energy degrees of freedom on the trimers
are denoted using the operators $\h{c}_{i\sigma}$, $c_{i\sigma}$ and
the classical $\rm Cr$ spins are denoted as $\vec{S}_a$. The trimer sublattice
is afforded hoppings $t$ and $t'$ which originate from direct overlaps between trimers (or indirectly via $\rm Cr$ or $\rm As$)
in the $ab$ planes and along the $c$ axis respectively.
Here we denote sites on the $\rm Fe$ sublattice as
$i$ and $j$, while sites on the $\rm Cr$ sublattice are
denoted as $a$ and $b$. The notation $a\in \hexagon_i$ indicates
that the sum on the $\rm Cr$ sublattice is over sites
in the hexagon that surround the trimer at $i$ on the $\rm Fe$ sublattice.
Furthermore $\avg{ij} \in ab,c$ denotes bonds in the $ab$ or $c$ planes
respectively. The interactions are given as follows: $U$ is the 
intra-trimer repulsion, inherited from the $\rm Fe$ atoms, $J_K$ is
the $\rm Fe-Cr$ exchange and $J_H$ is the $\rm Cr-Cr$ exchange. In this Hamiltonian 
the $\rm Cr$ spins appear as an effective magnetic field for the
trimers, which we will denote as $\vec{h}_i = \sum_{a \in \hexagon_i} \vec{S}_a$.

In the large $U$ limit, where $U \gg t$ and $t^2/U,(t')^2/U \ll J_K,J_H$, this model should reproduce
the model studied in Ref [\onlinecite{hopkinson}], and thus their classical results 
provide a useful point of comparison for our large $U/t$ behaviour. To attack the intermediate
$U/t$ regime we will use a slave-rotor approach, reviewed in the following section.

\section{Slave Rotors}
\label{slave-rotors}
For completeness and standardization of notation, we
review the general properties of two dimensional rotors and follow with a discussion of
the slave particle representation that bears their name\cite{rotor-florens-mott}.

A rotor in two dimensions is an object 
that possesses only angular momentum, prototypically of the form $H \propto L^2$ where $L$ 
is the angular momentum operator about some axis, say the $\hat{z}$ axis.
We classify states by the eigenstates of the $L$ operator, $L \ket{n} = n \ket{n}$ 
where $n$ is an integer. Raising and lowering operators are defined as
$$
  \h{U} \ket{n}  =  \ket{n+1},   \hspace{30pt}  U  \ket{n}  =  \ket{n-1},
$$
where $U$ is a unitary operator. From this definition it is simple 
to show that $L$ and $U$ satisfy the commutation relations,
$$
  \left[ L,U \right]      =  U,   \hspace{30pt} \left[ L,\h{U} \right]  =  -\h{U}.
$$
Since $U$ is unitary
it can be written as $U = \exp(-i\theta)$ where $\h{\theta} = \theta$ and one can show that this implies the canonical commutation relation $[\theta,L] = i$,
 showing that $L$ and $\theta$ are canonically conjugate variables.

To use these rotors as a slave-particle we associate the local electron basis with 
the product of the states of slave fermion and the states of an $O(2)$ rotor,
\begin{eqnarray*}
  \ket{0}        & = & \ket{0}_f \ket{+1}_{\theta} ,\\
  \ket{\up}      & = & \ket{\up}_f\ket{0}_{\theta} ,\\
  \ket{\down}    & = & \ket{\down}_f\ket{0}_{\theta} ,  \\
  \ket{\up\down} & = & \ket{\up \down}_f\ket{-1}_{\theta}.
\end{eqnarray*}
The slave fermion is called a spinon and spinon states are denoted by
an $f$ subscript. The slave-rotor will be referred to as a rotor and a $\theta$ subscript will be used to denote 
rotor states. The natural interpretation is to have the spinon to be neutral and the rotor to carry the charge of the electron,
thus explicitly separating the spin and charge degrees of freedom.
Having expanded our local Hilbert space, a constraint is required to remove the unphysical states. 
The four physical states above are characterized by
$$
L_i + \sum_{\sigma} n^f_{i\sigma} = 1,
$$
where $L_i$ is the rotor angular momentum operator and $n^f_{i\sigma}$ is the spinon number operator.
This is the Hilbert space constraint.
The electron operators can therefore be expressed as
$$
c_{i \sigma} = f_{i \sigma} e^{i \theta_i} ,
$$
where $\exp\left(-i \theta_i\right)$ is the rotor lowering operator and $f_{i\sigma}$, $\h{f}_{i\sigma}$ are
the fermionic spinon operators. Using this
representation the electronic Hamiltonian for the trimers (\ref{full-model}) is written as
\begin{eqnarray*}
H  & = &
  - t\sum_{\avg{ij}\in ab}\sum_\sigma \h{f}_{i\sigma}f_{j\sigma} e^{-i (\theta_i -\theta_j)} \\
& & - t'\sum_{\avg{ij}\in c}\sum_\sigma  \h{f}_{i\sigma}f_{j\sigma} e^{-i (\theta_i -\theta_j)} \\
& &+ \frac{U}{2} \sum_i L_i (L_i -1)+ \frac{J_K}{2} \sum_{i,  a\in \hexagon_i} \left(\h{f}_i\vec{\sigma}f_i \right)\cdot \vec{S}_a.
\end{eqnarray*}
The Hubbard term is now a kinetic term for the rotors, so the
complexity of the Hubbard interaction has been moved to the hopping
term and the constraint. We note that the $\rm Fe-Cr$ coupling term
only involves the spinon degrees of freedom.
\section{Mean Field Theory}
\label{mean-field-theory}
We approach this problem using mean field theory. For simplicity we take
the perspective of Florens and Georges,\cite{rotor-florens-mott} first decoupling
the hopping term into
$$
\h{f}_{i\sigma}f_{j\sigma} e^{-i (\theta_i -\theta_j)} \approx
\chi_{ij} e^{-i (\theta_i - \theta_j)} +
B_{ij} \h{f}_{i\sigma}f_{j\sigma} - \chi_{ij} B_{ij},
$$
where we have introduced the mean fields $\chi_{ij} = \frac{1}{2} \sum_{\sigma}\avg{\h{f}_{i\sigma}f_{j\sigma}}$ and
$B_{ij} =  \avg{e^{-i (\theta_i-\theta_j)}}$. Note that we have assumed that
$\chi_{ij}$ is independent of spin. To handle the constraint
we treat it both on average in space and on average in our states.
For the case of half-filling this leads to the two conditions,
$$
  \sum_i \avg{L_i}  =  0,  \hspace{30pt}  \sum_{i\sigma} \avg{n^{f}_{i\sigma}}  =  1. 
$$
Note that applying this on average
in space prohibits us from considering charge ordering in
our calculations. To enforce these constraints we introduce 
chemical potentials for the rotors and for the spinons, $\mu_L$ and $\mu_f$
respectively. This leads to two independent Hamiltonians which only talk to
each other through the mean fields $\chi_{ij}$ and $B_{ij}$,
\begin{eqnarray}
\label{spinon-ham}
H_{f} & = & - \sum_{\avg{ij}\sigma} (t_{ij}B_{ij} + \mu_f \delta_{ij}) \h{f}_{i\sigma}f_{j\sigma} \\
&&+ \frac{J_K}{2} \sum_{i,  a\in \hexagon_i} \left(\h{f}_i\vec{\sigma}f_i \right)\cdot \vec{S}_a,\\
\label{rotor-ham}
H_{L} & = & -2t \sum_{\avg{ij}} \chi_{ij} e^{-i \theta_i} e^{i \theta_j} + \frac{U}{2} \sum_i( L_i^2 - \mu_L L_i),
\end{eqnarray}
where we've introduced $t_{ij}$ which is equal to $t$ on the
in-plane triangular bonds and equal to $t'$ on the out of plane bonds.
While the spinon Hamiltonian can be treated using mean field theory,
the rotor Hamiltonian needs a different strategy. 

Two approaches for the rotor Hamiltonian have been used in the literature: 
a self-consistent cluster approach\cite{rotor-zhao} and a bosonic approach\cite{rotor-florens-mott}.
taking the bosonic approach, which is
most simply tackled using a path-integral formulation, the imaginary time action
takes the form,
\begin{eqnarray*}
 \nonumber
S(\theta,L) = 
\int_0^{\beta} & d\tau & \Big[ \sum_i \left(i  L_i \del_\tau \theta_i + \frac{U}{2} L_i^2 \right) \\
&-&  2\sum_{ij} t_{ij} \chi_{ij} e^{-i\theta_i} e^{i \theta_j} 
\Big],
\end{eqnarray*}
where $\mu_L$ is chosen to eliminate the linear terms in $S$.
Due to the symmetry of the action this guarantees
that $\avg{L_i}=0$. Next, we integrate out $L$ to get the following action,
\begin{equation}
\label{quantum-xy}
S(\theta) = 
\int_0^{\beta}  d\tau \left[  \frac{1}{2U}(\del_\tau\theta_i)^2- 2 \sum_{ij} t_{ij} \chi_{ij} e^{-i\theta_i} e^{i \theta_j} \right].
\end{equation}
We write this using a bosonic variable $\phi_i = e^{i\theta_i}$ 
subject to the constraint that $|\phi_i|^2 = 1$, giving
\begin{eqnarray}
\nonumber
\label{boson-action}
\nonumber
S(\cb{\phi},\phi;\lambda) & = &\int_0^{\beta} d\tau \Big[  \frac{1}{2U}\sum_i \left|\del_\tau \phi_i \right|^2
- i \sum_i \lambda_i \\
+ & &\sum_{ij}\left(i\lambda_i \delta_{ij} - 2 t_{ij} \chi_{ij}\right) \cb{\phi}_i \phi_j
\Big],
\end{eqnarray}
where $\lambda$ is an auxiliary field introduced to enforce the constraint.
Treating this new constraint in saddle point approximation,
the solution of the bosonic part of the Hamiltonian is reduced to solving
this saddle point equation and a free bosonic problem. These saddle point
equations simply fix the boson number at each site to one. 

\begin{figure}
\includegraphics{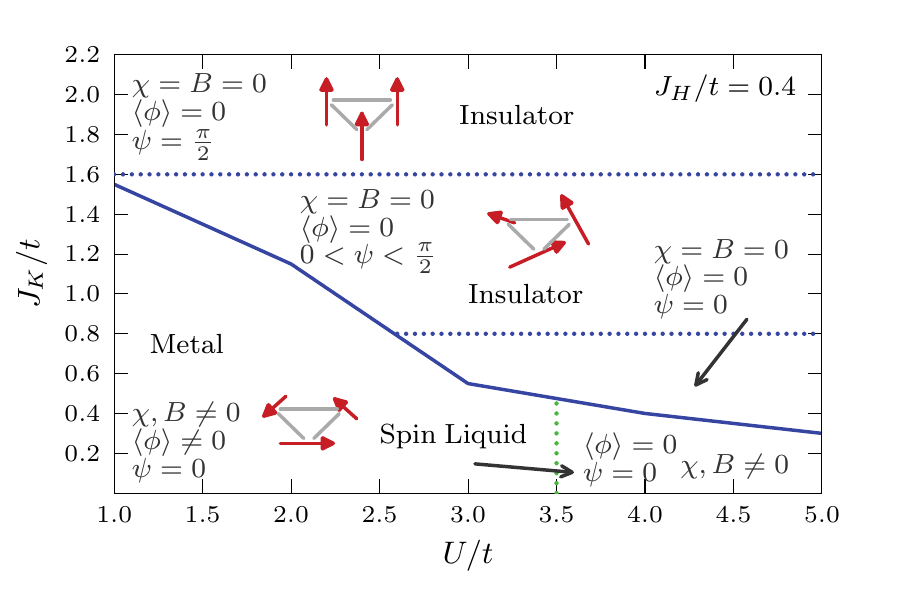}
\caption{ \label{phase4} (Color Online) Phase diagram as a function of $J_K/t$ and $U/t$ for
the value $J_H/t = 0.4$. The inset diagrams show a sample of the $\rm Cr$ 
spin configuration on one of the Kagome triangles.
$\chi$ and $B$ and the slave-rotor mean field parameters, $\avg{\phi}$ is the rotor
condensate and $\psi$ is the $\rm Cr$ canting angle (see text).
}
\end{figure}

We assume a uniform
state on the $\rm Fe$ sublattice, by considering $\chi_{ij} \equiv \chi$ and $B_{ij} \equiv B$
in plane, with $\chi_{ij} = \alpha \chi$ and $B_{ij} = \alpha B$ out of plane, where
$\alpha$ is chosen so that we smoothly match the non-interacting limit. 
\footnote{More general ansatzes could be employed while maintaining uniformity,
such as varying the phases of $\chi$ and $B$ over the tripled unit cell or 
include spin-dependent $\chi$. For simplicity we leave these considerations 
for future work.}
For the $\rm Cr$ spins it is natural to assume that the periodicity is that found by experiments\cite{julian,julian-neutron}
and previous classical calculations\cite{hopkinson}, with wavevector
of $(1/3,1/3,0)$. Within this space of magnetic states, we consider only canted
classical ground states, that is states where the inplane components in each
triangle are at $120^{\circ}$ but they are tilted out of plane by a canting angle $\psi$. These
states interpolate between a subset of the ground states for $J_H \gg J_K$ at $\psi = 0$
and the ferrimagnetic state valid for $J_K \gg J_H$ with $\psi = \pi/2$\cite{hopkinson}.
Among these states one can see that only those with a finite moment, as one sums
around a hexagon in the Kagome lattice, will be favoured due to the $J_K$ interaction.
With these constraints we only have a single set of magnetic states to consider
parametrized by the canting angle $\psi$. Under these assumptions the trimer feels the following effective magnetic field (defined
as $\vec{h}_i = \sum_{a \in \hexagon_i} \vec{S}_a$) after summing around the hexagon
\begin{equation*}
  \vec{h}_i = 6 S \cos{\psi}\left[ \cos{(\vec{Q}\cdot \vec{r}_i)}\hat{x}+
 \sin{(\vec{Q}\cdot \vec{r}_i)}\hat{y}\right]+
12 S \sin{\psi} \hat{z},
\end{equation*}
where $\vec{Q} = (1/3,1/3,0)$ and $S$ is the magnitude of the $\rm Cr$ moment.
Note the factor of $2$ between the in-plane and out-plane 
components due to the partial cancellation as we sum around the hexagon.
The phase diagram for $J_H/t = 0.4$ 
is shown in Fig. \ref{phase4}, and demonstrates the full range of phases.

There are three distinct phases shown in these phase diagrams. At low $J_K/t$ and $U/t$
we find a metallic phase on the $\rm Fe$ sublattice with zero canting angle (i.e. in plane, 120$^\circ$ ordering) on the $\rm Cr$ sublattice.
This metallic state is characterized by condensation of
the bosonic degree of freedom ($\avg{\phi} \neq 0$) and a uniform, real $\chi_{ij} = |\chi|$ 
implying the existence of an electron Fermi surface and gapless charge excitations.
Futhermore, the $\rm Fe$ trimers have an induced moment (anti-ferromagnetically) following the $\rm Cr$ moment. This moment
scales with $J_K/t$, and is thus will be small so long as $J_K/t$ is. As we increase $U/t$ and keep $J_K/t$ small, we find a metal insulator
transition near $U_c/t \approx 3.5$ into a uniform, $U(1)$ spin liquid (SL) phase. The metal-insulator boundary is very flat, as seen in Fig. \ref{phase4}.
, since under the mean field ansatz we have employed the critical $U$ is only a function of $\chi$ which changes only
by a small amount as we increase $J_K$ (below the jump into the paramagnet phase).
This SL phase is also characterized 
by a uniform, real $\chi_{ij}$ but gapped bosons ($\avg{\phi}=0$), meaning
the existence of a spinon Fermi surface but gapped charge excitations. This insulating phase carries
the induced moment as in the metallic phase, with the magnitude also being proportional to $J_K/t$. In both the metallic
and insulating phases as $J_K/t$ is increased (but remains small) the $\chi$ order parameter decreases towards zero. 

From the spin only model of Ref. [\onlinecite{hopkinson}] (accounting for the 
differing normalizations) one
expects that for large $U/t$ the canting angle will become non-zero at $J_K = 2 J_H$ and saturate at $J_K = 4 J_H$
as one sees in Fig. \ref{phase4}. We find that the uniform spin liquid does not support a non-zero canting angle. By this 
we mean that as $J_K/t$ is increased the canting angle remains zero until some critical $J_K/t$, wherein
the spin liquid is replaced by the canted AF with $\chi=B=0$. The converse 
is not true, as can be seen in Fig. \ref{phase4}. We see that the spin liquid phase 
can be destroyed before the onset of the canted magnetic state, leaving a window where we have a a simple insulating,
in-plane antiferromagnet. Once inside this canted AF phase with $\chi=B=0$, the canting angle increases with $J_K/t$ until in becomes saturated and we enter a ferrimagnetic phase, with
a net magnetic moment. The key feature of the phase diagrams we want to emphasize is that for a variety of values of $J_H$ and
for small $J_K$ this spin liquid state is stable and does not coexist with a finite canting.
\section{Discussion}
\label{discussion}
To discuss the applications of this to $\FeCrAs$ the effects of fluctuations must be taken into account
near the metal insulator transtion. These include both charge and gauge degrees of freedom and are treated
in detail in the work of Podolsky et al\cite{podolsky}. In this work it is found that there are two
relevant temperature scales $\cc{T}$ and ${T}^{**}$ that determine the qualitative features of the thermodynamic
and transport properties. These temperature scales vanish as one approaches the critical point
separating the metal and insulator. At temperatures above these scales the specific heat has weak logarithmic corrections
\begin{equation}
  C \sim T \ln\ln(1/T),
\end{equation}
while the conductivity has a strong temperature dependence. Specifically, writing $\sigma = \sigma_f + \sigma_b$ where $\sigma_f$
is the spinon conductivity and $\sigma_b$ is boson conductivity, under the assumption of weak disorder we have $\sigma_f \sim \sigma^{\rm imp}_f$
due to impurity scattering and
\begin{equation}
  \sigma_b \sim T \ln^2(1/T).
\end{equation}
This implies that the effects of the fluctuations on the specific heat are much more difficult to discern experimentally
then the effect on the resistivity, which will be a monotonically decreasing function of $T$, as seen in the experiments
on $\FeCrAs$. This is possible scenario for $\FeCrAs$ where a nearly linear specific is observed, as these logarithmic corrections
would only be visible over large temperature ranges, where other contributions would begin to dominate and wash out the signature.
This is also consistent with the magneto-resistance measurements\cite{julian}, where no change is seen in the low temperature
resistivity under magnetic fields up to $8$T, as the magnetic field would only couple weakly to the rotor fluctuations.

A method to test this hypothesis experimentally would be a study of the pressure dependence
of the transport and thermodynamic properties. Naively one expects that the application of
pressure should drive the material through the metal-insulator transition, into the quantum critical metal
and eventually into a Fermi liquid phase. Under our scenario, this could in principle be visible as
the development of a maximum in the resistivity at low temperatures as the pressure is increased. 
Furthermore, the specific heat should be relatively unaffected, still only being renormalized by a
logarithmic term.

The limitations of the current study deserve some discussion as
they lead to future directions. Due to the natural ansatz 
used in the slave-rotor study, a number of non-trivial spin-liquid states have been
excluded from the analysis, such as those with a non-trivial phase structure or
that break translational symmetry\cite{rau-prl}. A full exploration of the possible
phases in this model could yield useful insights for $\FeCrAs$. Another aspect of this problem that requires future
work is the inclusion quantum effects in the description of $\rm Cr$, leading to a
Kondo-Heisenberg model with strong interactions for the conduction electrons. While
the addition of frustrating interactions for the Kondo spins has attracted attention
recently\cite{coleman1,si1}, the inclusion of electronic interactions is largely unaddressed (particularly
with frustration on the conduction electrons) and is an interesting, but highly non-trivial,
question for future study.

\section{Summary}
\label{conclusions}
In this paper we have presented and motivated a minimal model
for the low energy degrees of freedom in the compound $\FeCrAs$. 
Starting
from the crystal structure and using the experimental facts, we have
argued that the magnetic degrees of freedom are well described by
a set of classical, localized moments and the electronic degrees
of freedom take the form of a half-filled Hubbard model on the
trimer sublattice. The coupling between these two subsystems stabilizes
a definite magnetic order on the localized moments despite the high degree
frustration. To explain the thermodynamic and transport properties of this
material at low temperatures we propose that the electrons residing
on the $\rm Fe$ trimers could be close to a quantum critical point separating metallic
and insulating phase. The charge fluctuations associated with the critical
point strongly renormalize the transport properties but provide only small corrections
to the thermodynamics, qualitatively consistent the experimental results
on $\FeCrAs$. Finally, we have discussed unexplored experimental consequences of this proposal
and future directions for theoretical work.

\begin{acknowledgments}
We thank J. Hopkinson, S. Julian, Y.B. Kim and Y.J. Kim for useful discussions.
This research was supported by NSERC of Canada and the Canada Research Chair program.
\end{acknowledgments}

\appendix

\section{Hopping integrals}
\begin{figure}
\vspace{10pt}
\subfigure[\ $xy$ orbital]{\includegraphics[width=60pt]{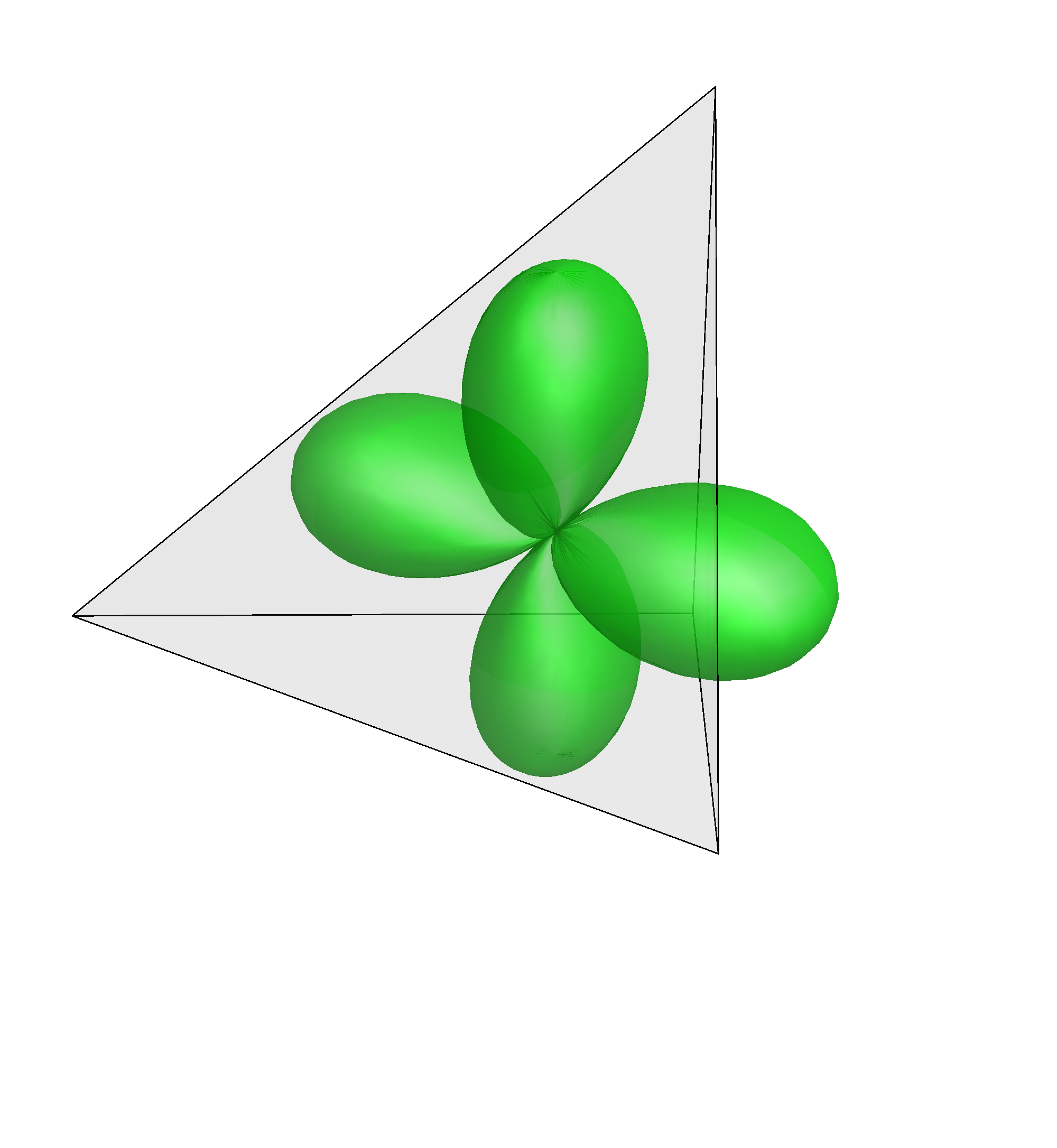}}
\subfigure[\ $xz$ orbital]{\includegraphics[width=60pt]{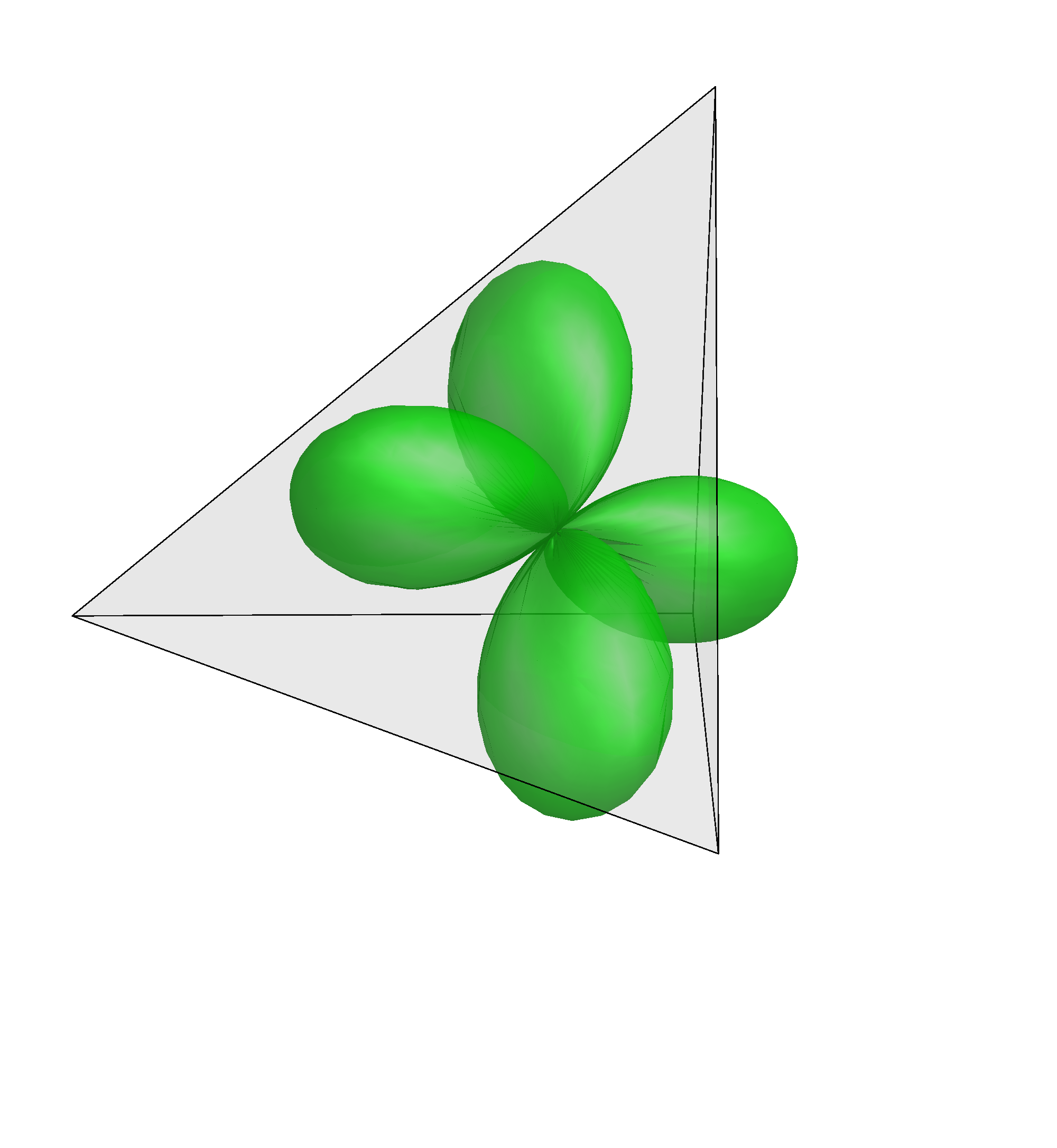}}
\subfigure[\ $yz$ orbital]{\includegraphics[width=60pt]{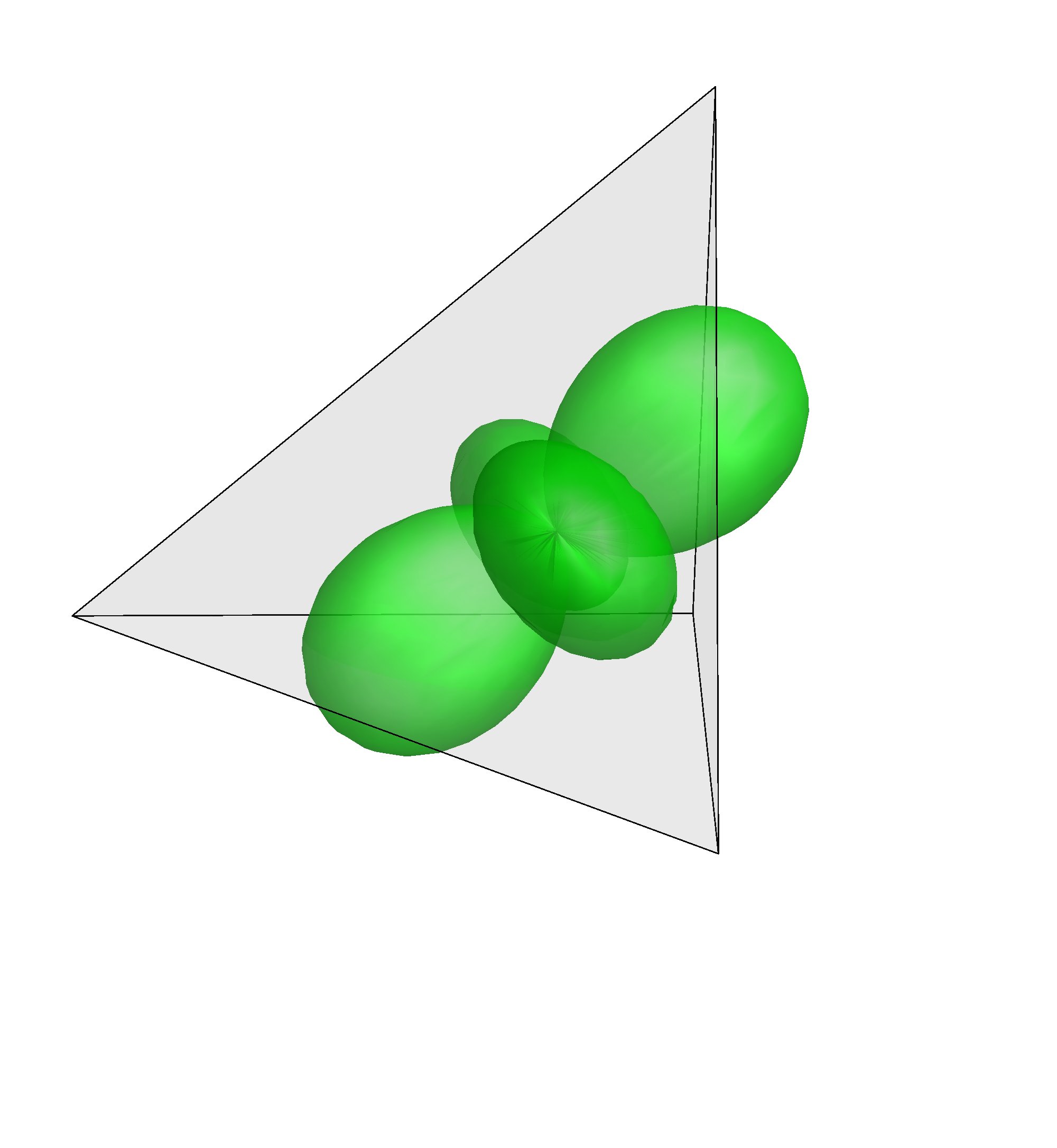}}
\caption{(Color Online) The orbitals of the $t_2$ level, rotated along the local axes of
an As tetrahedron.\label{local-orb}}
\end{figure}

We first consider direct hopping between the Fe atoms in a trimer. 
The $t_2$ level is composed of $xz$, $yz$ and $xy$ orbitals with respect to the canonical choice
of tetrahedron axes. Rotating these into the axes of a tetrahedron in a trimer, the orbitals are oriented
as shown in Fig. \ref{local-orb}.

From Fig. \ref{local-orb} one can estimate that
the only significant hoppings would be between ${xy-xy}$, ${xz-yz}$ and ${yz-xz}$. One can approach this
more quantitatively using the ideas of Slater-Koster\cite{slater-koster} theory to compute
the orbital overlaps in terms of rotation matrices and irreducible overlaps. We identify the vector connecting two sites in a trimer $\hat{a}$ as $(\hat{x}+\sqrt{3}\hat{y})/2$, giving 
the (Euler) representation $R=R_{\hat{z}}(0) R_{\hat{y}}(-\frac{\pi}{2}) R_{\hat{z}} (-\frac{\pi}{3})$. The rotation that takes our tetrahedron into the proper axes is given
as $R_T=R_{\hat{z}}(-\frac{\pi}{2}) R_{\hat{y}}(\frac{\pi}{2}) R_{\hat{z}} (\frac{\pi}{4})$ giving the transformation to local axes $R_T$ and $R_{\hat{z}}(\frac{2\pi}{3}) R_T$ for the neighbouring tetrahedron in the trimer.
This reduces the number of parameters to three, given by overlaps of $l=2$, $m=0,\pm 1,\pm 2$ orbitals displaced
along the $\hat{z}$ direction which we will denote as $t_{\sigma}$, $t_{\pi}$ and $t_{\delta}$ respectively. The hopping matrix in the basis of $xz$,$yz$ and $xy$ is
then given by
\begin{equation}
\nonumber
\frac{1}{32}
\left[
\begin{tabular}{ccc}
 $\scriptstyle t_{\delta} -8t_{\pi} - 9t_{\sigma}$ &
 $\scriptstyle 7t_{\delta} -16t_{\pi} + 9t_{\sigma}$ &
 $\scriptstyle \sqrt{\frac{3}{2}}\left(-7t_{\delta} -4t_{\pi} +3t_{\sigma}\right)$ \\
 $\scriptstyle 7t_{\delta} -16t_{\pi} + 9t_{\sigma}$ &
 $\scriptstyle t_{\delta} -8t_{\pi} - 9t_{\sigma}$ &
 $\scriptstyle \sqrt{\frac{3}{2}}\left(-7t_{\delta} +4t_{\pi} -3t_{\sigma}\right)$ \\
 $\scriptstyle \sqrt{\frac{3}{2}}\left(7t_{\delta} +4t_{\pi} -3t_{\sigma}\right)$ &
 $\scriptstyle \sqrt{\frac{3}{2}}\left(-7t_{\delta} -4t_{\pi} +3t_{\sigma}\right)$ &
 $\scriptstyle \frac{1}{2}\left(49t_{\delta} -12t_{\pi} +3t_{\sigma}\right)$  \\
\end{tabular}
\right].
\end{equation}
Considering the orbitals at atomic separations, we expect $t_{\sigma}$ and $t_{\delta}$ are positive with
$t_{\pi}$ negative. The simplest ansatz to try is $ t_{\sigma} = t_{\delta} = -t_{\pi}\equiv  t$. This gives:

\begin{equation}
t
\left(
\begin{tabular}{ccc}
 $0$ & $1$ & $0$ \\
 $1$ & $0$ & $0$ \\
 $0$ & $0$ & $1$ 
\end{tabular}
\right),
\end{equation}
as one might guess by looking at the orbital overlaps in the rotated axes.
Varying the numerical values for 
the irreducible hopping parameters around this point gives qualitatively the same picture
as this simple case, justifying our naive guess.

\bibliography{fecras-paper}

\end{document}